\documentclass[journal]{IEEEtran}

\usepackage[cmex10]{amsmath}
\usepackage{array}
\usepackage{url}
\usepackage{amsfonts}
\usepackage{amssymb}
\usepackage{graphicx}

\begin{document}

\newtheorem{corollary}{Corollary}
\newtheorem{definition}{Definition}
\newtheorem{example}{Example}
\newtheorem{lemma}{Lemma}
\newtheorem{proposition}{Proposition}
\newtheorem{theorem}{Theorem}
\newtheorem{fact}{Fact}
\newtheorem{property}{Property}
\newcommand{\bra}[1]{\langle #1|}
\newcommand{\ket}[1]{|#1\rangle}
\newcommand{\braket}[3]{\langle #1|#2|#3\rangle}
\newcommand{\ip}[2]{\langle #1|#2\rangle}
\newcommand{\op}[2]{|#1\rangle \langle #2|}

\newcommand{\tr}{{\rm tr}}
\newcommand{\supp}{{\it supp}}
\newcommand{\sch}{{\it Sch}}

\newcommand {\E } {{\mathcal{E}}}
\newcommand {\F } {{\mathcal{F}}}
\newcommand {\diag } {{\rm diag}}

\title{Distinguishability of Quantum States by Separable Operations\thanks{This work was partly supported
by the Natural Science Foundation of China (Grant Nos. 60621062 and
60503001) and the Hi-Tech Research and Development Program of China
(863 project) (Grant No. 2006AA01Z102).}}
\author{Runyao Duan,\ \ Yuan Feng,\ \,\ \ Yu Xin, \ \ and Mingsheng Ying\thanks{The material in this paper was presented in part as a long talk
at the 2007 Asia Conference of Quantum Information Science (AQIS07),
Kyoto, Japan. The authors are with the State Key Laboratory of
Intelligent Technology and Systems, Tsinghua National Laboratory for
Information Science and Technology, Department of Computer Science
and Technology, Tsinghua University, Beijing 100084, China. Yu Xin
is also with the Department of Physics, Tsinghua University, Beijing
100084, China. E-mails: dry@tsinghua.edu.cn (Runyao Duan),
feng-y@tsinghua.edu.cn (Yuan Feng), xiny05@mails.tsinghua.edu.cn (Yu
Xin), and yingmsh@tsinghua.edu.cn (Mingsheng Ying)}}
\date{\today}
\maketitle

\begin{abstract}
We study the distinguishability of multipartite quantum states by
separable operations. We first  present a necessary and sufficient
condition for a finite set of orthogonal quantum states to be
distinguishable by separable operations. An analytical version of
this condition is derived for the case of $(D-1)$ pure states, where
$D$ is the total dimension of the state space under consideration. A
number of interesting consequences of this result are then carefully
investigated. Remarkably, we show there exists a large class of
$2\otimes 2$ separable operations not being realizable by local
operations and classical communication. Before our work only a class
of $3\otimes 3$ nonlocal separable operations was known [Bennett et
al, Phys. Rev. A \textbf{59}, 1070 (1999)]. We also show that any
basis of the orthogonal complement of a multipartite pure state is
indistinguishable by separable operations if and only if this state
cannot be a superposition of $1$ or $2$ orthogonal product states,
i.e., has an orthogonal Schmidt number not less than $3$, thus
generalize the recent work about indistinguishable bipartite
subspaces [Watrous, Phys. Rev. Lett. \textbf{95}, 080505 (2005)].
Notably, we obtain an explicit construction of indistinguishable
subspaces of dimension $7$ (or $6$) by considering a composite
quantum system consisting of two qutrits (resp. three qubits), which
is slightly better than the previously known indistinguishable
bipartite subspace with dimension $8$.

\smallskip\
\end{abstract}

\section{Introduction}
One of the most profound features of quantum mechanics is that
composite quantum systems can exhibit nonlocality. Such an effect
can be interpreted as there exist some global quantum operations
performing on a composite system cannot be implemented by the owners
of the subsystems using local operations and classical communication
(LOCC) only. Actually, it is well known that any locally realizable
quantum operation is necessarily separable. The converse part,
however, is not always true, as a consequence of the weird
phenomenon of ``nonlocality without entanglement" discovered by
Bennet and coworkers \cite{BDF+99}. On the one hand, although many
partial progresses have been made, the structure of LOCC operations
is far from well understood. On the other hand, the class of
separable operations is rather restricted and is with rich
mathematical structure. It is relatively easier to determine whether
a given quantum operation is separable by employing the well
developed tools for the separability of quantum states. Thus, a deep
understanding of separable operations is of particular importance in
quantum information theory.

A general strategy  for studying quantum nonlocality is to consider
what kind of information processing tasks can be achieved by LOCC.
Roughly speaking, if a certain task is accomplished with different
optimal global and local efficiencies, then we can construct a class
of quantum operations that cannot be realized by LOCC. Among these
tasks, discrimination of orthogonal quantum states is a very
effective one and has been received considerable attentions in
recent years. Many interesting results have been reported, see Refs.
\cite{BDF+99,BDM+99,DMS+00,WSHV00,VSPM01, CY02,
HM03,JCY05,ABB+05,OGA06,WH02, TDL01, DLT02,
EW02,CHE04,BW06,NAT05,OH06, GKR+01,FAN04,WAT05,HK05,
HMM+06,DFJY07,HSSH03,CL03,COH07, FS07} and references therein for a
partial list. There exist sets of orthogonal product states that
cannot be discriminated perfectly by LOCC \cite{BDF+99,BDM+99,
DMS+00}; a perfect discrimination between two multipartite
orthogonal quantum states can always be achieved locally
\cite{WSHV00}, and this  result can be generalized to the case when
distinguishing two nonorthogonal states or two infinite dimensional
orthogonal states \cite{VSPM01,CY02,HM03,JCY05,ABB+05,OGA06}; a
complete characterization for the local distinguishability of
$2\otimes 2$ states also has been obtained \cite{WH02}. Despite of
these significant advances, determining the local distinguishability
of a finite set of orthogonal states on a multipartite state space
still is a formidable task. Actually, even for an arbitrary set of
multipartite orthogonal product states we still don't know how to
decide the local distinguishability analytically except for the
special cases of $2\otimes n$ and $3\otimes 3$, which were solved in
Ref. \cite{BDM+99} and Ref. \cite{FS07}, respectively.

Very recently separable operations have been widely used in order to
obtain useful criteria for local distinguishability. In particular,
Chefles employed separable operations as a tool to obtain a
necessary and sufficient condition for a set of general quantum
states to be probabilistically distinguishable by LOCC \cite{CHE04},
see also Ref. \cite{BW06}. Nathanson showed that the number of
$d\otimes d$ maximally entangled states distinguishable by separable
operations is at most $d$ \cite{NAT05} and the same result was
independently obtained by Owari and Hayashi using a slightly
different method \cite{OH06}, which extensively generalized the
results obtained by Ghosh {\it et al.} \cite{GKR+01} and by Fan
\cite{FAN04}. Watrous found a bipartite subspace with interesting
property that no bases distinguishable by separable operations, and
employed it to explicitly construct a class of quantum channels with
suboptimal environment-assisted capacity \cite{WAT05}, which settled
an open problem suggested by Hayden and King \cite{HK05}. Hayashi
{\it et al.} obtained a connection between the number of
distinguishable states by separable operations and the average
entanglement degree of the quantum states to be discriminated
\cite{HMM+06}. In Ref. \cite{DFJY07} we studied the local
distinguishability of an arbitrary basis of a multipartite state
space and provided a universal tight lower bound on the number of
locally unambiguously distinguishable members in an arbitrary basis.
All these results suggest the following question:``What kind of
quantum states can be perfectly distinguishable by separable
operations?" The answer to this question will not only lead to a
better understanding of the nature of separable operations, but also
accelerate the present research of quantum nonlocality.

The purpose of this paper is to study the strength and the
limitation of separable operations by considering the
distinguishability of quantum states under separable operations. We
present a necessary and sufficient condition for the
distinguishability of a set of general multipartite quantum states.
Assisting with this condition, we completely solve the problem of
distinguishing $(D-1)$ multipartite pure states by separable
operations, where $D$ is the dimension of the multipartite state
space under consideration. Two consequences are of particular
interests. First we show that there exists a large class of
separable operations performing on a $2\otimes 2$ quantum system
cannot be realized by LOCC. This is somewhat surprising as it
indicates that the capability of separable operations is much more
powerful than that of LOCC operations even when only the simplest
composite quantum system is under consideration, which settles an
open problem recently posed by Gheorghiu and Griffiths \cite{GG07}
and highlights the nonlocal nature of separable operations. It is
also worth noting that before our work only nonlocal separable
operations acting on $3\otimes 3$ systems are known \cite{BDF+99}.

Second we show that any basis of the orthogonal complement of a
multipartite pure state is indistinguishable by separable operations
if and only if this state cannot be a superposition of $1$ or $2$
orthogonal product states, i.e., has an orthogonal Schmidt number
not less than $3$, thus provide a generalization of the result by
Watrous \cite{WAT05}, who proved that any basis of the orthogonal
complement of a $d\otimes d$ maximally entangled state is
indistinguishable by separable operations when $d\geq 3$.
Furthermore,  we also present an explicit construction of
indistinguishable bipartite (or tripartite) subspaces of dimension
$7$ (resp. $6$), which reduces the minimal dimension of
indistinguishable bipartite subspace from $8$ (given by Watrous in
Ref. \cite{WAT05}) to $7$.

Throughout this paper we consider a multipartite quantum system
consisting of $K$ parts, say $A_1,\cdots, A_K$. We assume part $A_k$
has a state space $\mathcal{H}_k$ with dimension $d_k$. The whole
state space is given by $\mathcal{H}=\otimes_{k=1}^K\mathcal{H}_k$
with total dimension $D=d_1\cdots d_K$. Sometimes we use the
notation $d_1\otimes \cdots \otimes d_K$ for $\mathcal{H}$. With
these assumptions, the rest of the paper is organized as follows. In
Section II we give some necessary preliminaries, including a brief
review of the notion of separable operations, the concept of
generalized Schmidt number of a quantum state, and some technical
lemmas. In Section III we present a necessary and sufficient
condition for the distinguishability of a set of orthogonal quantum
states on $\mathcal{H}$ by separable operations (Theorem
\ref{exactdis}). Many immediate but interesting consequences of this
condition are discussed in details. Sections IV and V devote to the
following specific problem: Let $\ket{\Phi}$ be a pure state on
$\mathcal{H}$ and let
$\mathcal{B}=\{\ket{\Psi_1},\cdots,\ket{\Psi_{D-1}}\}$ be an
orthonormal basis of $\{\ket{\Phi}\}^\perp$(i.e., the orthogonal
complement of $\ket{\Phi}$), determine the distinguishability of
$\mathcal{B}$ by separable operations. We show this problem can be
solved analytically. Two special cases are most notable. The first
case is concerned with two qubits, i.e., $K=2$ and $d_1=d_2=2$. We
find a necessary condition for the distinguishability of
$\{\ket{\Psi_1},\ket{\Psi_2},\ket{\Psi_3}\}$ by separable operations
is that the summation of the concurrences of $\ket{\Psi_k}$ is equal
to that of $\ket{\Phi}$ (Theorem \ref{2otimes2sep}). When
$\ket{\Phi}$ is maximally entangled, such a condition is also
sufficient (Corollary \ref{2ME}). The second case of interest is
that any basis $\mathcal{B}$ of $\{\ket{\Phi}\}^\perp$ is
indistinguishable by separable operations if and only if the
orthogonal Schmidt number of $\ket{\Phi}$ (the least number of
orthogonal product states that can linearly express $\ket{\Phi}$) is
not less than $3$ (Theorem \ref{indissubspace}). Section VI is of
independent interest. We give an explicit construction of an
indistinguishable subspace of dimension $7$ (or $6$) when
considering a composite quantum system consisting of two qutrits
(resp. three qubits). We conclude the paper in Section VII, and put
some additional proofs in Appendix.

\section{Preliminaries}  A general quantum state $\rho$ is
characterized by its density operator, which is a positive
semi-definite operator with trace one on $\mathcal{H}$. The density
operator of a pure state $\ket{\psi}$ is simply the projector
$\op{\psi}{\psi}$. The support of $\rho$, denoted by $\supp(\rho)$,
is the vector space spanned by the eigenvectors of $\rho$ with
positive eigenvalues. Alternatively, suppose $\rho$ can be
decomposed into a convex combination of $\op{\psi_k}{\psi_k}$, say,
\begin{equation}\label{essemble}
\rho=\sum_{k=1}^n p_k\op{\psi_k}{\psi_k},
\end{equation}
where $0<p_k\leq 1$ and $\sum_{k=1}^n p_k=1$. Then
$\supp(\rho)=span\{\ket{\psi_k}:1\leq k\leq n\}$. This fact will be
extensively used without being explicitly stated. In particular,
$\rho$ is said to be separable if $\ket{\psi_k}$ can be chosen as
product states.

A nonzero positive semi-definite operator $E$ on $\mathcal{H}$ is
said to be separable if the normalized form ${E}/{\tr(E)}$ is a
separable quantum mixed state. A separable positive operator-valued
measure (POVM) is  a collection of semi-definite positive operators
$\Pi=\{\Pi_1,\cdots, \Pi_n\}$ such that $\sum_{k=1}^n
\Pi_k=I_{\mathcal{H}}$ and $\Pi_k$ is separable for each $1\leq
k\leq n$, where $I_{\mathcal{H}}$ is the identity operator on
$\mathcal{H}$. Now we are ready to introduce the notion of separable
operation. Intuitively, a separable operation is a trace-preserving
completely positive linear map which sends separable states to
separable ones. For completeness, we state a formal definition as
follows.
\begin{definition}\label{so-def}\upshape
A separable operation is a quantum operation with product Kraus
operators. More precisely, a separable operation $\E$ acting on a
multipartite state space $\mathcal{H}=\otimes_{j=1}^K\mathcal{H}_j$
should be of the following form:
\begin{equation}
\E(\rho)=\sum_{k=1}^N E_k\rho E_k^\dagger,
\end{equation}
where $E_k=\otimes_{j=1}^K E_{kj}$ satisfies $\sum_{k=1}^N
E_{k}^\dagger E_k=I$, and $E_{kj}$ is linear operator on
$\mathcal{H}_j$.
\end{definition}
The set of separable operations is a rather restricted class of
quantum operations. It is not difficult to show that any LOCC
operation is separable, see Ref. \cite{BDF+99} for a detailed
discussion. However, the converse part is not true as there exists
some separable operation acting on $3\otimes 3$ cannot be
implemented locally \cite{BDF+99}.

Let $\mathcal{S}=\{\rho_1,\cdots, \rho_n\}$ be a collection of $n$
quantum states. We say that $\mathcal{S}$ is perfectly
distinguishable by separable measurements if there is a separable
POVM $\Pi$ such that $\tr(\Pi_k\rho_j)=\delta_{k,j}$ for any $1\leq
k,j\leq n$. Interestingly, if a set of quantum states
$\{\rho_1,\cdots, \rho_n\}$ can be perfectly distinguishable by a
separable POVM $\Pi$, then we can transform this set of quantum
states into another set of LOCC distinguishable states by applying a
suitable separable operation $\E$ on the quantum system. More
precisely, let $\{\sigma_1,\cdots, \sigma_n\}$ be any set of
separable quantum states  that can be perfectly distinguishable by
LOCC, then the mentioned separable operation can be constructed as
follows:
\begin{equation}\label{separableoperation}
\mathcal{E}(\rho)=\sum_{k=1}^n \tr(\Pi_k\rho)\sigma_k.
\end{equation}
One can easily verify that $\mathcal{E}(\rho_k)=\sigma_k$. The
function of $\mathcal{E}$ can be intuitively understood as
performing a separable POVM $\{\Pi_k\}$ on $\rho$ and then preparing
a separable state $\sigma_k$ according to the outcome $k$, finally
forgetting the classical information. Clearly, $\mathcal{E}$ is a
separable quantum operation and can be used to perfectly
discriminate the given states. Due to the above reason, when a set
of states are perfectly distinguishable by a separable POVM, we
directly say they are perfectly distinguishable by separable
operations.

The rest of this section is to present some useful notations and
technical lemmas.
\begin{lemma}\label{tech}\upshape
Let $E$ be a positive semi-definite operator such that  $0\leq E\leq
I_{\mathcal{H}}$, and let $\rho$ be a density matrix on
$\mathcal{H}$. Then $\tr(E\rho)=1$ if and only if $E-P\geq 0$, where
$P$ is the projector on the support of $\rho$.
\end{lemma}

\textit{Proof.} We first show that if $E-P\geq 0$ and $0\leq E\leq
I$, then $\tr(E\rho)=1$. To see this, let $\ket{\psi}$ be any
normalized vector in $\supp(\rho)$. Then $\braket{\psi}{P}{\psi}=1$.
It follows from $E-P\geq 0$ that $\braket{\psi}{E}{\psi}\geq 1$. On
the other hand, $E\leq I$ implies $\braket{\psi}{E}{\psi}\leq 1$. So
for any $\ket{\psi}\in \supp(\rho)$ we have
$$\tr(E\op{\psi}{\psi})=\braket{\psi}{E}{\psi}=1.$$
Noticing that $\rho$ can be decomposed into $\sum_{k=1}^n p_k
\op{\psi_k}{\psi_k}$ such that $\ket{\psi_k}\in \supp(\rho)$ and
$\sum_{k=1}^n p_k=1$, we have
$$\tr(E\rho)=\sum_{k=1}^n p_k\tr(E\op{\psi_k}{\psi_k})=1.$$

Conversely, $\tr(E\rho)=1$ and $\braket{\psi}{E}{\psi}\leq 1$ imply
that $\tr(E\op{\psi}{\psi})=1$ for any normalized vector
$\ket{\psi}\in \supp(\rho)$. So $E\ket{\psi}=\ket{\psi}$ or
$E\op{\psi}{\psi}=\op{\psi}{\psi}$. Noticing that $P=\sum_{k=1}^n
\op{\psi_k}{\psi_k}$ such that $\{\ket{\psi_k}\}$ is an orthonormal
basis for $\supp(\rho)$, then $EP=P$ and $PE=P$. Let $Q=I-P$. Then
$$E=(P+Q)E(P+Q)=P+QEQ,$$
which implies $E-P=QEQ\geq 0$.\hfill $\blacksquare$

We also need the following concept of Schmidt number of a quantum
state.

\begin{definition}\label{sch-def}\upshape
The {\it Schmidt number} of a multipartite quantum state $\rho$,
denoted by $\sch(\rho)$, is the minimal number of product pure
states whose linear span contains $\supp(\rho)$. The {\it orthogonal
Schmidt number} of $\rho$, denoted by $\sch_{\perp}(\rho)$, can be
defined similarly with the additional requirement that these product
pure states should be mutually orthogonal.
\end{definition}

For pure state $\ket{\Psi}$ the (orthogonal) Schmidt number is
exactly reduced to the minimal number of (orthogonal) pure product
states needed to express $\ket{\Psi}$ as a superposition and this is
just the definition given by Eisert and Briegel \cite{EB01}. It is
difficult to determine the Schmidt number of a general multipartite
state. Only for bipartite pure states and some special multi-qubit
states the Schmidt number can be analytically determined \cite{EB01,
AACJ+00}. We should point out that the definition of Schmidt number
given here does not coincide with the one given by Terhal and
Horodecki \cite{TH00}.

By definition,  we have the following useful fact:
\begin{equation}\label{lowerbound}
\sch(\rho)\geq \max\{\sch(\Psi): \ket{\Psi}\in \supp(\rho)\}.
\end{equation}
This inequality can be used to provide a lower bound for
$\sch(\rho)$.

It is interesting that the Schmidt number provides a simple
criterion for the separability. In fact, when $\rho$ is separable,
$supp(\rho)$ can be spanned by a set of product states. Thus we
have:
\begin{lemma}\label{schmidtsep}\upshape
For any quantum state $\rho$, $R(\rho)\leq \sch(\rho)$, where
$R(\rho)$ is the rank of $\rho$. The equality holds if $\rho$ is
separable.
\end{lemma}

A typical use of Lemma \ref{schmidtsep} is to show a quantum state
$\rho$ is entangled. Usually this can be achieved by finding a
vector $\ket{\Psi}$ in $\supp(\rho)$ such that $\sch(\Psi)>R(\rho)$.

The Schmidt decomposition of a bipartite pure state is not unique.
However, when multipartite pure states with Schmidt number $2$ are
under consideration, we do have a unique decomposition. Let
$\ket{a}=\otimes_{k=1}^K\ket{a_k}$ and
$\ket{b}=\otimes_{k=1}^K\ket{b_k}$ be product vectors on
$\mathcal{H}=\otimes_{k=1}^K\mathcal{H}_k$.  Then $\ket{a}$ and
$\ket{b}$ are said to be different at the $k$-th entry if
$\ket{a_k}\neq z\ket{b_k}$ for any complex number $z$. Let
$h(\ket{a},\ket{b})$ be the total number of different entries
between $\ket{a}$ and $\ket{b}$. Then we have the following
interesting result.
\begin{lemma}\label{uniquedecom}\upshape
Let $\ket{\Phi}=\ket{a}+\ket{b}$ be an entangled state of
$\mathcal{H}$ such that $h(\ket{a},\ket{b})\geq 3$. Then there
cannot exist product vectors $\ket{c}$ and $\ket{d}$ such that
$\ket{\Phi}=\ket{c}+\ket{d}$ and $\{\ket{a},\ket{b}\}\neq
\{\ket{c},\ket{d}\}$. In other words, $\ket{\Phi}$ has a unique
Schmidt decomposition.
\end{lemma}

\textit{Proof.} Without loss of generality, assume
$h(\ket{a},\ket{b})=3$. By contradiction, suppose there exist
another two product vectors $\ket{c}$ and $\ket{d}$ which also yield
a decomposition of $\ket{\Phi}$. That is,
\begin{equation}\label{schmidtdecom}
\ket{a_1a_2a_3}+\ket{b_1b_2b_3}=\ket{c_1c_2c_3}+\ket{d_1d_2d_3},
\end{equation}
where $a_k$ is not equal to $b_k$ for each $1\leq k\leq 3$. Let
$\ket{a_1^\perp}\in span\{\ket{a_1},\ket{b_1}\}$ such that
$\ip{a_1^\perp}{a_1}=0$ and $\ip{a_1^\perp}{b_1}\neq 0$. Then
multiplying $\bra{a_1^\perp}$ on the both sides of the above
equation we have that $\ket{b_2b_3}$ is contained in
$S=span\{\ket{c_2c_3},\ket{d_2d_3}\}$. Similarly we can also show
$\ket{a_2a_2}\in S$. On the other hand, there are only two unique
product vectors $\ket{c_2c_3}$ and $\ket{d_2d_3}$ in $S$ (up to some
factors). Then it follows that $\{\ket{a_2a_3}, \ket{b_2b_3}\}$ is
essentially $\{\ket{c_2c_3}, \ket{d_2d_3}\}$. Without loss of
generality, assume $\ket{a_2a_3}=\alpha \ket{c_2c_3}$ and
$\ket{b_2b_3}=\beta\ket{d_2d_3}$ for some complex numbers $\alpha$
and $\beta$, then
\begin{equation}\label{contra}
(\alpha\ket{a_1}-\ket{c_1})\ket{c_2c_3}+(\beta\ket{b_1}-\ket{d_1})\ket{d_2d_3}=0,
\end{equation}
which is possible if and only if $\ket{a_1}=\alpha^{-1}\ket{c_1}$
and $\ket{b_1}=\beta^{-1}\ket{d_1}$. That also means
$\ket{a}=\ket{c}$ and $\ket{b}=\ket{d}$. A contradiction with our
assumption.\hfill $\blacksquare$

We also need the following result concerning with the separability
of rank $2$ positive semi-definite operators.

\begin{lemma}\label{2sumsep}\upshape
Let $\ket{\Psi}$ and $\ket{\Phi}$ be pure states on $\mathcal{H}$
and $\lambda\geq 0$. Then
$\rho(\lambda)=\op{\Psi}{\Psi}+\lambda\op{\Phi}{\Phi}$ is separable
if and only if one of the following cases holds:

(i) Both $\ket{\Psi}$ and $\ket{\Phi}$ are product states and
$\lambda\geq 0$;

(ii) $\ket{\Psi}$ is a product state and $\lambda=0$;

(iii) There are two product states $\ket{a}$ and $\ket{b}$ and
positive real numbers $\alpha$, $\beta$, $\gamma$, $\delta$ such
that $\ket{\Psi}=\alpha\ket{a}+\beta\ket{b}$,
$\ket{\Phi}=\gamma\ket{a}-\delta\ket{b}$, and
$\lambda=\alpha\beta/\gamma\delta$. (There may be some global phase
factors before $\ket{\Psi}$ and $\ket{\Phi}$.)
\end{lemma}

\textit{Proof.} By a direct calculation one can readily verify that
(i), (ii) and (iii) are sufficient for the separability of
$\rho(\lambda)$. We only need to show that they are also necessary.
The case that both states are unentangled is trivial, corresponding
to (i). We shall consider the following two cases only:

(1) One of $\ket{\Psi}$ and $\ket{\Phi}$ is entangled. First assume
that $\ket{\Phi}$ is entangled. Then for any $\lambda>0$.
$\rho(\lambda)$ is just a mixture of an entangled state and a
product state, and should be entangled for any $\lambda>0$, as shown
by Horodecki {\it et al.} \cite{HSTT03}. So the only possible case
is $\lambda=0$, i.e., (ii) holds. Second,  when $\ket{\Psi}$ is
entangled we can show by a similar argument that $\rho(\lambda)$
cannot be separable for any $\lambda\geq 0$.

(2) Both states are entangled. We shall show that condition (iii) is
necessary. Suppose that $\rho(\lambda)$ is separable, then by Lemma
\ref{schmidtsep}, it follows that
$\sch(\rho(\lambda))=R(\rho(\lambda))=2$. In other words, the
support of $\rho(\lambda)$ can be spanned by two product states
$\ket{a}$ and $\ket{b}$. Noticing that both $\ket{\Psi}$ and
$\ket{\Phi}$ are contained in the support of $\rho(\lambda)$, we can
write
\begin{eqnarray}
\ket{\Psi}&=&\alpha\ket{a}+\beta\ket{b},\\
\ket{\Phi}&=&\gamma\ket{a}+\delta\ket{b}
\end{eqnarray}
for nonzero complex numbers $\alpha,\beta,\gamma, \delta$.
Furthermore, $\ket{a}$ and $\ket{b}$ are also the only product
states in $\supp(\rho(\lambda))$. So $\rho(\lambda)$ should be a
mixture of $\op{a}{a}$ and $\op{b}{b}$. Substituting $\ket{\Psi}$
and $\ket{\Phi}$ into $\rho(\lambda)$ and setting the coefficients
of $\op{a}{b}$ and $\op{b}{a}$ to be  zero we have
\begin{equation}
\alpha\beta^*+\lambda\gamma\delta^*=0.
\end{equation}
By adding suitable phase factors before $\ket{\Psi}$ and
$\ket{\Phi}$, we can make $\alpha$ and $\gamma$ positive. The
condition $\lambda>0$ implies that $\beta=|\beta|e^{i\theta}$ and
$\delta=-|\delta|e^{i\theta}$ for some real number $\theta$.
Absorbing the phase factor $e^{i\theta}$ into $\ket{b}$ and
resetting $\beta$ and $\delta$ to $|\beta|$ and $-|\delta|$
respectively, we know that condition (iii) is satisfied. \hfill
$\blacksquare$
\section{Conditions for the distinguishability of quantum states by separable
operations}It would be desirable to know when a collection of
quantum states is perfectly distinguishable by separable operations.
Orthogonality is generally not sufficient for  the existence of a
separable POVM for discrimination. A rather simple but useful
necessary and sufficient condition is as follows.
\begin{theorem}\label{exactdis}\upshape
Let $\mathcal{S}=\{\rho_1,\cdots,\rho_n\}$ be a collection of
orthogonal quantum states of $\mathcal{H}$. Then $\mathcal{S}$ is
perfectly distinguishable by separable operations if and only if
there exist $n$ positive semi-definite operators $\{E_1,\cdots,
E_n\}$ such that $P_k+E_k$ is separable for each $1\leq k\leq n$,
and $\sum_{k=1}^n E_k=P_0$, where $P_k$ is the projector on
$\supp(\rho_k)$, and $P_0=I_{\mathcal{H}}-\sum_{k=1}^n P_k$.
\end{theorem}

\textit{Proof.} Sufficiency is obvious. Suppose that (ii) holds, let
us define a POVM $\Pi=\{\Pi_1,\cdots, \Pi_n\}$ as follows:
$\Pi_{k}=P_k+E_k$ for each $1\leq k\leq n$. It is easy to verify
that $\Pi$ is a separable measurement that perfectly discriminates
$\mathcal{S}$.

Now we turn to show  the necessity. Suppose $\mathcal{S}$ is
perfectly distinguishable by some separable POVM, say
$\{\Pi_1,\cdots,\Pi_n\}$. Take $E_k=\Pi_k-P_k$ for each $1\leq k\leq
n$. Then $\sum_{k=1}^n E_k=P_0$. To complete the proof, it suffices
to show $E_{k}\geq 0$. By the assumption, we have
$\tr(\Pi_k\rho_k)=1$.  Then the positivity of $E_k$ follows directly
from Lemma \ref{tech}. \hfill
$\blacksquare$\\

In some sense Theorem \ref{exactdis} is only a reformulation of our
discrimination problem. One may doubt it can provide any valuable
results. However, this seemingly trivial reformulation does give us
much more operational forms of the measurement operators for
discrimination and connects them to the separability problem, which
is one of the central topics in quantum information theory and has
been extensively studied since the last two decades. So many well
developed tools for the separability problem may be applicable in
the context of discrimination. As a result, Theorem \ref{exactdis}
is unexpectedly powerful.

Some special but interesting cases of Theorem \ref{exactdis} deserve
careful investigations. When the supports of the states in
$\mathcal{S}$ together span the whole state space, i.e.,
$supp(\sum_{k=1}^n\rho_k)=\mathcal{H}$, $\mathcal{S}$ is perfectly
distinguishable by separable operations if and only if $P_k$ is
separable for each $1\leq k\leq n$. In particular, an orthonormal
basis of $\mathcal{H}$ is perfectly distinguishable by separable
operations if and only if it is a product basis, which is a well
known result first obtained by Horodecki \textit{et al.} using a
rather different method \cite{HSSH03}, see also \cite{CHE04} for
another simple proof.

A slightly more general case is when there exists $k$ such $P_0+P_k$
and $P_j$ are all separable for any $j\neq k$. In this case we say
that $\{P_j:j\neq k\}$ is completable \cite{NOTE1}. It has been
shown in Ref. \cite{BDM+99} that a sufficient condition for a set of
orthogonal product states to be completable is that they are exactly
distinguishable by LOCC. Combining this with Theorem \ref{exactdis},
we have the following interesting corollary:
\begin{corollary}\label{UPB-sep}\upshape
A set of LOCC distinguishable product states together with any state
in its orthogonal complement is always perfectly distinguishable by
separable operations.
\end{corollary}

We notice that DiVincenzo and co-workers have shown in Ref.
\cite{DMS+00} that any three (or four) orthogonal multipartite
(resp. bipartite) product states are distinguishable by LOCC. Hence
we conclude immediately by the above corollary that any four (or
five) orthogonal multipartite (resp. bipartite) states including
three (resp. four) product states are perfectly distinguishable by
separable operations. The case when $\mathcal{S}$ is a set of
product states is of particular interest and has been studied
carefully in Ref. \cite{DMS+00} using a rather different method.
Specifically, let
$\mathcal{S}=\{\ket{\psi_1},\cdots,\ket{\psi_n}\}$. Then it has been
shown that if $S_k=S-\{\ket{\psi_k}\}$ is completable for each
$1\leq k\leq n$, then $S$ is perfectly distinguishable by separable
operations. Interesting examples include any orthogonal UPB
consisting of four $2\otimes 2\otimes 2$ (or five $3\otimes 3$)
product states. Clearly we have refined the results by DiVincenzo
{\it et al.}

As another important consequence of Theorem \ref{exactdis}, we shall
show that the problem of distinguishing $(D-1)$ orthogonal pure
states by separable operations can be completely solved. Notice that
if $\ket{\Phi}$ is a product state, then the only form of
$\mathcal{B}$ that can be distinguishable by separable operations is
a product basis. For simplicity, in what follows we shall assume
$\ket{\Phi}$ is an entangled state. We shall consider two cases
$\sch_\perp(\Phi)=2$ and $\sch_\perp(\Phi)\geq 3$ respectively in
the next two sections.

\section{Distinguishability of arbitrary orthonormal basis of
$\{\ket{\Phi}\}^\perp$ with $\sch_\perp(\Phi)=2$}

We start with the simplest $2\otimes 2$ case and provide an
analytical solution. It is well known that any $2\otimes 2$ state
$\ket{\Psi}$ can be uniquely represented by a $2\times 2$ matrix
$\Psi$ as follows:
\begin{equation}\label{isomorphism}
\ket{\Psi}=(I\otimes \Psi)\ket{\Phi_+},
\end{equation}
where $\ket{\Phi_+}=1/\sqrt{2}(\ket{00}+\ket{11})$ is a maximally
entangled state. The \textit{concurrence} of $\ket{\Psi}$ is given
by the absolute value of the determinant of $\Psi$, i.e.,
$C(\Psi)=|\det(\Psi)|.$ This rewriting form  coincides with the
original definition given by Wootters \cite{WOO98}. So $0\leq
C(\Psi)\leq 1$ for any $2\otimes 2$ state $\ket{\Psi}$. In
particular, $C(\Psi)$ vanishes for product states while attains one
for maximally entangled states. We say complex numbers $z_1$ and
$z_2$ are anti-parallel if there exists a negative real number $a$
such that $z_1=az_2$. First we need the following simplified version
of Lemma \ref{2sumsep} in the $2\otimes 2$ case.
\begin{lemma}\label{2sep}\upshape
Let $\ket{\Psi}$ and $\ket{\Phi}$ be two $2\otimes 2$ entangled pure
states, and let $\lambda\geq 0$. Then
$\rho(\lambda)=\op{\Psi}{\Psi}+\lambda\op{\Phi}{\Phi}$ is separable
if and only if $\Psi\Phi^{-1}$ has two anti-parallel eigenvalues and
$\lambda=C(\Psi)/C(\Phi)$. Note that $\ket{\Phi}$ is entangled
implies that $\Phi$ is invertible.
\end{lemma}

Now the distinguishability of three $2\otimes 2$ orthogonal states
by separable operations is characterized  as follows.
\begin{theorem}\label{2otimes2sep}\upshape
Let $\ket{\Phi}$ be a $2\otimes 2$ pure state, and  let
$\mathcal{B}=\{\ket{\Psi_1},\ket{\Psi_2},\ket{\Psi_3}\}$ be an
orthonormal basis for $\{\ket{\Phi}\}^{\perp}$. Then $\mathcal{B}$
is perfectly distinguishable by separable operations if and only if
(i) $\Psi_k\Phi^{-1}$ has two anti-parallel eigenvalues for each
entangled state $\Psi_k$; and (ii)
$C(\Psi_1)+C(\Psi_2)+C(\Psi_3)=C(\Phi)$.
\end{theorem}

\textit{Proof.} By Theorem \ref{exactdis}, the POVM element that can
perfectly identify $\ket{\Psi_k}$ should have the form
$\Pi_k=\op{\Psi_k}{\Psi_k}+\lambda_k\op{\Phi}{\Phi},$ where $0\leq
\lambda_k\leq 1$ and $\lambda_1+\lambda_2+\lambda_3=1$. If
$\ket{\Psi_k}$ is a product state, then it follows from Lemma
\ref{2sumsep} that $\Pi_k$ is separable if and only if
$\lambda_k=0$. Otherwise by Lemma \ref{2sep} we have that $\Pi_k$ is
separable if and only if ${\Psi_k}{\Phi}^{-1}$ has two anti-parallel
eigenvalues and $\lambda_k=C(\Psi_k)/C(\Phi)$. So by
$\lambda_1+\lambda_2+\lambda_3=1$ we have
$C(\Psi_1)+C(\Psi_2)+C(\Psi_3)=C(\Phi)$. That completes the
proof.\hfill $\blacksquare$

The most interesting part in the above theorem is Condition (ii). It
reveals a precise relation between the concurrences and the
distinguishability by separable operations. By a careful
observation, we notice that a similar condition has been previously
obtained by Hayashi {et al.} in Ref. \cite{HMM+06}, where a very
general relation between distinguishability by separable operations
and average entanglement degree of the states to be discriminated
was presented. When only $2\otimes 2$ states were involved, this
relation can be reduced to an inequality similar to condition (ii).
The key difference is that here what we obtained is an exact
identity that (almost) completely characterizes the
distinguishability by separable operations.

Suppose now ${\ket{\Phi}}$ is maximally entangled. Then $\Phi=U$ for
some $2\times 2$ unitary matrix. Noticing that
$$\tr(\Psi_k\Phi^{-1})=\tr(U^\dagger {\Psi_k})=2\ip{\Phi}{\Psi_k}=0$$
for $k=1,2,3$, we conclude that $\Psi_k\Phi^{-1}$ should have two
anti-parallel eigenvalues (assume that $\ket{\Psi_k}$ is entangled).
Furthermore, the concurrence of a $2\otimes 2$ maximally entangled
state is one. Collecting all these facts together we have the
following interesting corollary:
\begin{corollary}\label{2ME}\upshape
Let $\ket{\Phi}$ be a $2\otimes 2$ maximally entangled state, and
let $\mathcal{B}=\{\ket{\Psi_1},\ket{\Psi_2},\ket{\Psi_3}\}$ be an
orthonormal basis for $\{\ket{\Phi}\}^{\perp}$. Then $\mathcal{B}$
is perfectly distinguishable by separable operations if and only if
$C(\Psi_1)+C(\Psi_2)+C(\Psi_3)=1$.
\end{corollary}

Theorem \ref{2otimes2sep} implies that there exists a large class of
$2\otimes 2$ separable operations that cannot be implemented
locally. We shall exhibit an explicit construction of such separable
operations. We achieve this goal by identifying a set of states that
is distinguishable by separable operations but is not LOCC
distinguishable.

Let $\ket{\Psi(\theta)}=\cos\theta\ket{01}+\sin\theta\ket{10}$ and
$\ket{\Phi(\theta)}=\cos\theta\ket{00}+\sin\theta\ket{11}$ for
$\theta\in\mathcal{R}$. Then for any $0<\alpha\leq \beta\leq \pi/4$,
let $\ket{\Phi}=\ket{\Phi(\beta)}$, and let
$\ket{\Psi_1}=\ket{\Psi(\alpha)},$
$\ket{\Psi_2}=\cos\gamma\ket{\Psi(\alpha-\frac{\pi}{2})}+\sin\gamma\ket{\Phi(\beta-\frac{\pi}{2})}$,
$\ket{\Psi_3}=\sin\gamma\ket{\Psi(\alpha-\frac{\pi}{2})}-\cos\gamma\ket{\Phi(\beta-\frac{\pi}{2})}$
be an orthonormal basis of $\{\ket{\Phi}\}^\perp$. Choose $\gamma$
such that
\begin{equation}\label{gamma}
\tan^{-1}\sqrt{\frac{\sin 2\alpha}{\sin 2\beta}}\leq \gamma\leq
\tan^{-1} \sqrt{\frac{\sin 2\beta}{\sin 2\alpha}}.
\end{equation}
By direct calculations we can easily  verify that the assumptions of
Theorem \ref{2otimes2sep} are fulfilled, thus
$\mathcal{B}=\{\ket{\Psi_1},\ket{\Psi_2},\ket{\Psi_3}\}$ is
perfectly distinguishable using separable operations. However, it is
clear whenever $0<\alpha<\beta\leq \frac{\pi}{4}$, both
$\ket{\Psi_1}$ and $\ket{\Psi_2}$ are entangled, thus $\mathcal{B}$
is indistinguishable by LOCC, as shown by Walgate and Hardy
\cite{WH02}. Interestingly, $\ket{\Psi_3}$ is reduced to a product
state when $\gamma$ takes one of the extreme values in Eq.
(\ref{gamma}).

The above example also implies that the distinguishability by
separable operations has some continuous property. We formalize this
intuitive observation as follows:
\begin{theorem}\label{continuous}\upshape
Let $c_1,c_2,c_3$ be nonnegative real numbers such that $0\leq
c_1+c_2+c_3\leq 1$. Then there exists a set of three $2\otimes 2$
states $\{\ket{\Psi_1},\ket{\Psi_2},\ket{\Psi_3}\}$ satisfying (i)
$C(\Psi_k)=c_k$ for $k=1,2,3$; and (ii)
$\{\ket{\Psi_1},\ket{\Psi_2},\ket{\Psi_3}\}$ is perfectly
distinguishable by separable operations.
\end{theorem}

There is a nice geometrical way to illustrate  the
distinguishability of quantum states by different class of
operations. To see this, we identify a triple of orthogonal quantum
states $(\ket{\psi_1},\ket{\psi_2},\ket{\psi_3})$ by a point
$(C(\psi_1),C(\psi_2),C(\psi_3))$ of $R^{3}$. We say such a point is
distinguishable by global (or separable, LOCC) operations if the
related states are distinguishable by the corresponding operations.
By a careful analysis, we can show any point in the regular
tetrahedron in Fig. 1 corresponds to some orthonormal basis of
$\{\ket{\Phi_+}\}^{\perp}$. A detailed proof for this interesting
fact is given in the Appendix A. So the points in the regular
tetrahedron represent a region that can be perfectly distinguishable
by global quantum operations. The yellow triangle  $BCD$ is exactly
the set of points distinguishable by separable operations. Finally,
three vertices $B$, $C$, and $D$ are the only points that are
distinguishable by LOCC. From such a representation we can clearly
see that the class of $2\otimes 2$ separable operations strictly
lies between the classes of LOCC and global operations.

\begin{figure}[ht]
  \centering
  \includegraphics[scale=0.7]{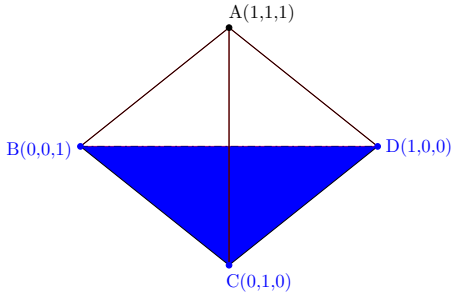}
  \caption{Illustration of the distinguishability of the basis of $\{\ket{\Phi_+}\}^\perp$, where $\ket{\Phi_+}$ is a $2\otimes 2$
  maximally entangled states and we identify an orthonormal basis $(\ket{\psi_1},\ket{\psi_2},\ket{\psi_3})$
  by a point $(C(\psi_1),C(\psi_2),C(\psi_3))$ in $R^3$. Any point in the regular tetrahedron ABCD corresponds a legal orthonormal basis
  of $\{\ket{\Phi_+}\}^\perp$, thus is distinguishable by global operations. Vertices B, C, and D are the only points corresponding
  to LOCC distinguishable basis. Interestingly, the triangle $\bigtriangleup BCD$ represents bases that are distinguishable by separable operations. }
  \label{2}
\end{figure}

Now we consider the multipartite setting. There are two possible
cases: $\ket{\Phi}$ is reduced to a bipartite entangled state only
between two parties; or $\ket{\Phi}$ is an entangled state at least
between three parties. The following two theorems deal with these
two cases separately. Surprisingly, the first case is essentially
reduced to $2\otimes 2$ case.
\begin{theorem}\label{schmidt2}\upshape
Let $\ket{\Phi}=\ket{a_1\cdots a_{K-2}}\ket{\Phi'}$ be a pure state
on $\mathcal{H}$ such that ${\ket{\Phi'}}$ is an entangled state
between parties $A_{K-1}$ and $A_{K}$. Then $\mathcal{B}$ is
perfectly distinguishable by separable operations if and only if
each entangled state $\ket{\Psi}$ in $\mathcal{B}$ can be written
into the form $\ket{\Psi}=\ket{a_1\cdots a_{K-2}}\ket{\Psi'}$ for
some bipartite entangled state $\ket{\Psi'}$ between $A_{K-1}$ and
$A_{K}$ such that (i) all ${\ket{\Psi'}}$ and ${\ket{\Phi'}}$ can be
embeded into a $2\otimes 2$ subspace $\mathcal{H}'$ of
$\mathcal{H}_{K-1}\otimes \mathcal{H}_{K}$; and (ii)
$\Psi'{\Phi'}^{-1}$ has two anti-parallel eigenvalues and
$\sum_{\Psi'} C(\Psi')=C(\Phi')$, where $\Psi'$ and $\Phi'$ are
$2\otimes 2$ matrices that are the representations of $\ket{\Psi'}$
and $\ket{\Phi'}$ on $\mathcal{H}'$ in the sense of Eq.
(\ref{isomorphism}), respectively.
\end{theorem}

\textit{Proof.} Let $\ket{\Psi}$ be any entangled state of
$\mathcal{B}$. Then there is some $0<\lambda\leq 1$ such that
$\op{\Psi}{\Psi}+\lambda\op{\Phi}{\Phi}$ is separable. According to
the assumptions on $\ket{\Phi}$ and employing Lemma \ref{2sumsep},
we can easily see that $\ket{\Psi}$ should be of the form
$\ket{b_1\cdots b_{K-2}}\otimes \ket{\Psi'}$. Second, we have
$\ket{a_1\cdots a_{K-2}}$ should be identical to $\ket{b_1\cdots
b_{K-2}}$ up to some factors. Otherwise
$span\{\ket{\Psi},\ket{\Phi}\}$ cannot contain product states. So we
need $\op{\Psi'}{\Psi'}+\lambda\op{\Phi'}{\Phi'}$ being separable.
Now the problem is reduced to the $2\otimes 2$ case and the result
follows directly from Theorem \ref{2otimes2sep}. \hfill
$\blacksquare$

The case that $\ket{\Phi}$ is at least entangled among three parties
is much more simple. Any distinguishable basis should contain a
unique entangled state with a predetermined form.
\begin{theorem}\upshape
Let $\ket{\Phi}=\cos\theta\ket{a}+\sin\theta\ket{b}$ be an entangled
pure state of $\mathcal{H}$ such that $\ip{a}{b}=0$ and
$h(\ket{a},\ket{b})\geq 3$. Then $\mathcal{B}$ can be perfectly
distinguishable by separable operations if and only if there is a
unique entangled state $\ket{\Psi}$ in $\mathcal{B}$ with the form
$\ket{\Psi}=\sin\theta\ket{a}-\cos\theta\ket{b}$, and the other
states should form an orthogonal product basis of
$\{\ket{a},\ket{b}\}^\perp$.
\end{theorem}
\textit{Proof.} Sufficiency is obvious. We only consider the
necessity. Suppose that $\mathcal{B}$ is distinguishable by
separable operations. Noticing that $\ket{\Phi}$ is entangled, we
conclude that $\mathcal{B}$ should contain at least one entangled
state. By Lemma \ref{uniquedecom}, one can easily show that any
entangled state $\ket{\Psi}$ such that
$\op{\Psi}{\Psi}+\lambda\op{\Phi}{\Phi}$ is separable should  be
contained in $span\{\ket{a},\ket{b}\}$. Noticing that $\ket{\Psi}$
is orthogonal to $\ket{\Phi}$, we can deduce that $\ket{\Psi}$ is
uniquely determined and is given by
$\sin\theta\ket{a}-\cos\theta\ket{b}$. \hfill $\blacksquare$
\section{$\{\ket{\Phi}\}^\perp$ has no bases distinguishable by
separable operations when $\sch_\perp(\Phi)\geq 3$} Now we consider
the case when $\sch_\perp(\Phi)\geq 3$. Our result is a
generalization of Watrous's result. It is remarkable that the
orthogonal Schmidt number of $\ket{\Phi}$ completely characterizes
the distinguishability of the subspace $\{\ket{\Phi}\}^\perp$.

\begin{theorem}\label{indissubspace}\upshape
Let $\ket{\Phi}$ be an entangled pure state on $\mathcal{H}$. Then
$\{\ket{\Phi}\}^{\perp}$ having no orthonormal basis perfectly
distinguishable by separable operations if and only if
$Sch_{\perp}(\Phi)>2$. In particular, when $Sch_{\perp}(\Phi)=2$,
there always exists an orthonormal basis $\mathcal{B}$ of
$\{\ket{\Phi}\}^{\perp}$ that is perfectly distinguishable by LOCC.
\end{theorem}

\textit{Proof.} Necessity: Suppose $Sch_\perp(\Phi)=2$. Then the
existence of a distinguishable basis by separable operations follows
directly from Theorem \ref{schmidt2}. Here we further construct a
basis for $\{\ket{\Phi}\}^\perp$ that is distinguishable by LOCC.
Notice that $\ket{\Phi}$ can be written into the form
$\ket{\Phi}=\cos\theta\ket{\Phi_0}+\sin\theta\ket{\Phi_1},$ where
$\ket{\Phi_0}$ and $\ket{\Phi_1}$ are orthogonal product states on
$\mathcal{H}$ and $0<\theta<\pi/2$. Then we can extend
$\ket{\Phi_0}$ and $\ket{\Phi_1}$ into a complete orthogonal product
basis $\{\ket{\Phi_0},\ket{\Phi_1},\cdots, \ket{\Phi_{D-1}}\}$ of
$\mathcal{H}$. We can further assume this basis is distinguishable
by local projective measurements \cite{BDM+99}. Replacing
$\ket{\Phi_0}$ and $\ket{\Phi_1}$ with
$\ket{\Psi}=\sin\theta\ket{\Phi_0}-\cos\theta\ket{\Phi_1}$ we obtain
a basis for $\{\ket{\Phi}\}^\perp$ that is  distinguishable by the
same local projective measurements.

Sufficiency: In this case we have $Sch_{\perp}(\Phi)\geq 3$. If
$Sch(\Phi)\geq 3$. Then for any $\lambda_k>0$ we know from Lemma
\ref{2sumsep} that
$\Pi(\lambda_k)=\op{\Psi_k}{\Psi_k}+\lambda_k\op{\Phi}{\Phi}$ should
be entangled. That implies any basis of $\{\ket{\Phi}\}^\perp$ is
indistinguishable by separable operations.

Suppose now that $Sch_{\perp}(\Phi)\geq 3$ and $Sch(\Phi)=2$. By
contradiction, assume there exists a basis $\mathcal{B}$
distinguishable by separable operations. Then applying Lemma
\ref{uniquedecom}, we know there are two unique product vectors
$\ket{a}$ and $\ket{b}$ such that $\ket{\Phi}=\ket{a}+\ket{b}.$ So
any state $\ket{\Psi_k}$ with $\lambda_k>0$ should also be a
superposition of $\ket{a}$ and $\ket{b}$. As there are only two
orthogonal states in $span\{\ket{a},\ket{b}\}$, we conclude the only
possibility is that there is a unique state $\ket{\Psi_k}$ with
$\lambda_k=1$ and all the other states are product states. On the
other hand, let $\ket{\Psi}$ be the  entangled state in
$span\{\ket{a},\ket{b}\}$ such that $\ip{\Psi}{a}=0$. Then
\begin{equation}
\op{\Psi_k}{\Psi_k}+\op{\Phi}{\Phi}=\op{a}{a}+\op{\Psi}{\Psi}
\end{equation}
is separable. But the right hand side of the above equation is a
summation of an entangled state and a product state, it should be
entangled \cite{HSTT03}. A contradiction. \hfill $\blacksquare$

Let us check some interesting examples. Take $\ket{\Phi}$ to be a
$W$-type state, $\ket{\Phi}=a\ket{001}+b\ket{010}+c\ket{100},$ where
$|a|^2+|b|^2+|c|^2=1$, and $abc\neq 0$. Then
$Sch_\perp(\Phi)=Sch(\Phi)=3$. It follows from Theorem
\ref{indissubspace} that any orthonormal basis of
$\{\ket{\Phi}\}^{\perp}$ cannot be perfectly distinguishable by
separable operations. This yields an indistinguishable subspace with
dimension $2^3-1=7$, which is a slight improvement over the
bipartite case, where a $3\otimes 3$ indistinguishable subspace of
dimension $8$ was given by Watrous \cite{WAT05}. But if we choose
the $GHZ$-type state $\ket{\Phi}=\cos\theta
\ket{000}+\sin\theta\ket{111},$ where $0\leq \theta\leq
\frac{\pi}{2}$. Then $\{\ket{\Phi}\}^{\perp}$ does have an
orthonormal basis that is perfectly distinguishable by LOCC.
Interestingly, if we take
$\ket{\Phi}=\alpha\ket{000}+\beta\ket{+++},$ where $\alpha\beta\neq
0$ and $\ket{+}=(\ket{0}+\ket{1})/\sqrt{2}$. Then we can easily see
that $Sch(\Phi)=2$ but $\sch_\perp (\Phi)\geq 3$, as $\ket{\Phi}$
cannot be written into a superposition of two orthogonal product
states. Thus it follows from the above theorem that
$\{\ket{\Phi}\}^\perp$ does not have any orthonormal basis
distinguishable by separable operations.

\section{Indistinguishable subspaces with smaller dimensions}
The dimension of indistinguishable subspaces can be further reduced.
We shall show that there exist a $3\otimes 3$ indistinguishable
subspace with dimension $7$ and a $2\otimes 2\otimes 2$
indistinguishable subspace with dimension $6$. We first consider
bipartite case. Let
$\ket{\Phi_1}=(\ket{00}+\ket{11}+\ket{22})/\sqrt{3}$ and
$\ket{\Phi_2}=\ket{01}$. Let $S=span\{\ket{\Phi_1},\ket{\Phi_2}\}$.
It is clear that $S^{\perp}$ is a bipartite subspace with dimension
$7$. Surprisingly, we have the following interesting result:
\begin{theorem}\upshape\label{dim7}
$S^{\perp}$ is indistinguishable by separable operations.
\end{theorem}

\textit{Proof.} We need the following easily verifiable properties
of $S$ to complete the proof:

P0: $\ket{\Phi_2}$ is the unique product vector (up to a scalar
factor) in $S$;

P1. $\sch(\Psi)=3$ for any entangled state $\ket{\Psi}\in S$;

P2. Any positive operator $\rho$ such that $\supp(\rho)=S$ satisfies
$\sch(\rho)=4$.

The validity of P0 can be directly verified. We only consider P1 and
P2. We first show the validity of P1. Actually, any entangled state
$\ket{\psi}$ from $S$ can be written into a superposition of
$\ket{\Phi_1}$ and $\ket{\Phi_2}$. That is,
\begin{equation}
\ket{\psi}=\alpha\ket{\Phi_1}+\beta\ket{\Phi_2},
\end{equation}
where $\alpha\neq 0$ and $|\alpha|^2+|\beta|^2=1$. Substituting
$\ket{\Phi_1}$ and $\ket{\Phi_2}$ into the above equation, we have
\begin{equation}
\ket{\psi}=\frac{1}{\sqrt{3}}(\ket{0}(\alpha\ket{0}+\sqrt{3}\beta\ket{1})+\alpha\ket{11}+\alpha\ket{22}),
\end{equation}
which is clearly an entangled state with Schmidt number $3$ for any
$\alpha\neq 0$.

Now we turn to show the validity of P2. By contradiction, suppose
for some $\rho$ such that $\supp(\rho)=S$ we have $Sch(\rho)=3$.
Then there should exist three unnormalized product states
$\ket{a_1b_1}$,$\ket{a_2b_2}$, and $\ket{a_3b_3}$ such that
\begin{eqnarray}
\ket{\Phi_1}&=&\ket{a_1b_1}+\ket{a_2b_2}+\ket{a_3b_3},\\
\ket{\Phi_2}&=&\alpha_1\ket{a_1b_1}+\alpha_2\ket{a_2b_2}+\alpha_3\ket{a_3b_3}.
\end{eqnarray}
Without loss of generality, we may assume $\alpha_1\neq 0$. Then it
is clear that
\begin{equation}
\ket{\Phi_1}-\alpha_1^{-1}\ket{\Phi_2}=(1-\frac{\alpha_2}{\alpha_1})\ket{a_2b_2}+(1-\frac{\alpha_3}{\alpha_1})\ket{a_3b_3}.
\end{equation}
By P1, the left hand side of the above equation has Schmidt number
$3$, while the right hand side of the above equation has Schmidt
number at most $2$. That is a contradiction.

Now let $\mathcal{B}=\{\ket{\Psi_k}:1\leq k\leq 7\}$ be any
orthogonal basis of $S^\perp$. By Theorem \ref{exactdis}, the POVM
element identifying $\ket{\Psi_k}$ is of the form
$\Pi_k=\op{\Psi_k}{\Psi_k}+E_k$, where $\sum_{k=1}^7
E_k=\op{\Phi_1}{\Phi_1}+\op{\Phi_2}{\Phi_2}$ and $E_k\geq 0$.

So we can choose $E_k$ such that $E_k\neq 0$ and $E_k\neq
\op{01}{01}$ (up to some factor). We shall show that $\Pi_k$ should
be entangled. There are two cases. If $R(E_k)=1$ then by P1 we have
$\sch(E_k)=3$, which follows that $\sch(\Pi_k)\geq 3>R(\Pi_k)$.
Similarly, if $R(E_k)=2$ then by P2 we have $\sch(\Pi_k)\geq
\sch(E_k)=4>R(\Pi_k)$. In both cases $\Pi_k$ is entangled. That
completes the proof. \hfill $\blacksquare$

Using the very same method, we can construct a $2\otimes 2\otimes 2$
indistinguishable subspace with dimension $6$. For instance, take
$\ket{\Phi_1}=(\ket{001}+\ket{010}+\ket{100})/\sqrt{3}$ and
$\ket{\Phi_2}=\ket{000}$. Then $\{\ket{\Phi_1},\ket{\Phi_2}\}^\perp$
is indistinguishable by separable operations.

\section{Conclusions} In summary, we provided a necessary and
sufficient condition for the distinguishability of a set of
multipartite quantum states by separable operations. A set of three
$2\otimes 2$ pure states that is perfectly distinguishable by
separable operations but is indistinguishable by LOCC then was
explicitly constructed. As a consequence, there exists a large class
of nonlocal separable operations even for the simplest composite
quantum system consisting of two qubits. We also showed that the
orthogonal complement of a pure state has no bases distinguishable
by separable operations if and only if this state has an orthogonal
Schmidt number not less than $3$. We believe these results would be
useful in clarifying the relation between separable operations and
LOCC.

There are still a number of unsolved problems that are worthwhile
for further study. We have mentioned some  in the previous context.
Here we would like to stress two of them. The first one is
concerning with the distinguishability of orthogonal product pure
states. It a simple fact that any set of orthogonal product pure
states can be perfectly distinguishable by some
positive-partial-transpose preserving (PPT) operation. Is this  also
true for separable operations? We have seen that separable
operations are powerful enough to distinguish any set of orthogonal
product pure states in $3\otimes 3$ and $2\otimes 2\otimes 2$. If
the answer is affirmative in general, then how to prove? Otherwise a
counterexample would be highly desirable. With little effort we can
easily show that it is sufficient to verify whether any UPB is
perfectly distinguishable by separable operations. The difficulty we
met is that the structure of UPB on a multipartite state space
remains unknown except for some special cases.

Another problem is to find more applications of locally
indistinguishable subspaces (LIS). The work of Watrous suggests that
bipartite LIS can be used to construct quantum channels with
sub-optimal environment-assisted capacity. It would be of great
interest to employ LIS as a tool to give tighter upper of the
capacity. In the asymptotic setting similar problem has been
thoroughly studied by Winter \cite{WIN05}. Hopefully, these efforts
would reveal some deep properties of LIS and testify the richness of
the mathematical structure of this notion.
\section*{Appendix: Geometric representation of the concurrences of orthogonal bases of $\{\ket{\Phi}\}^\perp$}

First we recall a useful representation of the concurrence of a
two-qubit pure state. Let $\{\ket{\Phi_k}: 1\leq k\leq 4\}$ be the
magic basis \cite{WOO98}, and let $\ket{\psi}$ be any pure state
such that $\ket{\psi}=\sum_{k=1}^4 \lambda_k\ket{\Phi_k}$. Then the
concurrence of $\ket{\psi}$ is given by the following formula:
\begin{equation}\label{con-2}
C(\psi)=|\lambda_1^2+\lambda_2^2+\lambda_3^2+\lambda_4^2|.
\end{equation}
Suppose now we choose $\ket{\Phi}$ to be one of
$\{\ket{\Phi_k}:1\leq k\leq 4\}$, say $\ket{\Phi_4}$. Then any
vector from $\{\ket{\Phi}\}^\perp$ should be a linear combination of
$\{\ket{\Phi_1},\ket{\Phi_2},\ket{\Phi_3}\}$. As a direct
consequence, we have the following interesting lemma which connects
the concurrences of orthonormal bases of $\{\ket{\Phi}\}^\perp$ to
$3\times 3$ unitary matrices.

\begin{lemma}\label{con-unitary}\upshape
Let $\{\ket{\psi_1},\ket{\psi_2},\ket{\psi_3}\}$ be an orthogonal
basis for $\{\ket{\Phi}\}^\perp$, then there exists a $3\times 3$
unitary matrices $U=[u_{kl}]_{3\times 3}$ such that
\begin{eqnarray}\label{con-3}
C(\psi_1)=|u_{11}^2+u_{12}^2+u_{13}^2|,\\
C(\psi_2)=|u_{21}^2+u_{22}^2+u_{23}^2|,\\
C(\psi_3)=|u_{31}^2+u_{32}^2+u_{33}^2|.
\end{eqnarray}
Conversely, for any unitary matrix $U=[u_{kl}]_{3\times 3}$, there
exists an orthogonal basis for $\{\ket{\Phi}\}^\perp$, say
$\{\ket{\psi_1},\ket{\psi_2},\ket{\psi_3}\}$ such that Eq.
(\ref{con-3}) holds.
\end{lemma}
Motivated by the above lemma,  let $\mathcal{P}_1$ be the set of
$x\in \mathcal{R}^3$ such that there exists a unitary matrix
$U=[u_{kj}]_{3\times 3}$ such that
\begin{eqnarray}\label{P-1}
x_1&=&|u_{11}^2+u_{12}^2+u_{13}^2|,\\
x_2&=&|u_{21}^2+u_{22}^2+u_{23}^2|,\\
x_3&=&|u_{31}^2+u_{32}^2+u_{33}^2|.
\end{eqnarray}
Then it is clear that $\mathcal{P}_1$ is exactly the set of points
corresponding to the concurrences of the orthogonal bases of
$\{\ket{\Phi}\}^\perp$. Let $\mathcal{P}_2$ be the set of $x\in
\mathcal{R}^3$ satisfying the following equations:
\begin{eqnarray}\label{P-2}
x_1+x_2+x_3\geq 1,\\
x_1+x_2-x_3\leq 1,\\
x_2+x_3-x_1\leq 1,\\
x_3+x_1-x_2\leq 1,\\
0\leq x_1,x_2,x_3\leq 1.
\end{eqnarray}
Obviously, $\mathcal{P}_2$ is just a unit regular tetrahedron. We
shall show that $\mathcal{P}_2$ is contained by $\mathcal{P}_1$.
That means any point of the regular tetrahedron corresponds to some
orthonormal basis of $\{\ket{\Phi}\}^\perp$.
\begin{theorem}
$\mathcal{P}_2\subseteq \mathcal{P}_1$.
\end{theorem}

\textit{Proof.} For any point $x\in \mathcal{P}_2$, we shall
construct a $3\times 3$ unitary operation $U$ such that Eq.
(\ref{P-1}) holds. Without loss of generality, we may assume
$x_1\geq x_2\geq x_3\geq 0$. So Eq. (\ref{P-2}) is reduced to the
following equations:
\begin{eqnarray}\label{simP-2}
x_1+x_2+x_3\geq 1,\\
x_1+x_2-x_3\leq 1,\\
0\leq x_3\leq x_2\leq x_1\leq 1.
\end{eqnarray}
Construct a unitary matrix $U$ as follows:
$$U=\left(
      \begin{array}{ccc}
        u_{11} & u_{12} & u_{13} \\
        u_{21} & u_{22} & u_{23} \\
        u_{31} & u_{32} & u_{33} \\
      \end{array}
    \right).\left(
             \begin{array}{ccc}
               1 & 0 & 0 \\
               0 & e^{i\theta} & 0 \\
               0 & 0 & e^{i\theta} \\
             \end{array}
           \right),
$$
where $[u_{kl}]_{3\times 3}$ is a real orthogonal matrix, and $0\leq
\theta\leq \pi$. By a simple calculation, we have
$$U=\left(
      \begin{array}{ccc}
        u_{11} & u_{12}e^{i\theta} & u_{13}e^{i\theta} \\
        u_{21} & u_{22}e^{i\theta} & u_{23}e^{i\theta} \\
        u_{31} & u_{32}e^{i\theta} & u_{33}e^{i\theta} \\
      \end{array}
    \right).
$$
Our purpose is to choose suitable real numbers $u_{11}$, $u_{21}$,
$u_{31}$, and $\theta$ such that
\begin{eqnarray}
u_{11}^2+(1-u_{11}^2)e^{2i\theta}&=&x_1,\\
u_{21}^2+(1-u_{21}^2)e^{2i\theta}&=&x_2,\\
u_{31}^2+(1-u_{31}^2)e^{2i\theta}&=&x_3,\\
u_{11}^2+u_{21}^2+u_{13}^2&=&1.
\end{eqnarray}
From the first three equations, we have
\begin{equation}\label{u1k}
u_{1k}^2=\frac{\sin\theta\pm\sqrt{x_k^2-\cos^2\theta}}{2\sin\theta},~k=1,2,3.
\end{equation}
We shall choose a suitable $\theta$ such that the last normalized
relation holds.  That is, choose $\theta$ such that
$$\sin\theta\pm\sqrt{x_1^2-\cos^2\theta}\pm\sqrt{x_2^2-\cos^2\theta}\pm\sqrt{x_3^2-\cos^2\theta}=0,$$
where $\cos^{-1}(x_3)\leq \theta\leq \pi/2$. The difficulty here is
how to choose suitable signs for $u_{1k}$. Let
\begin{eqnarray}
f(\theta)=\sin\theta-\sqrt{x_1^2-\cos^2\theta}-\sqrt{x_2^2-\cos^2\theta}-\sqrt{x_3^2-\cos^2\theta},\nonumber\\
g(\theta)=\sin\theta-\sqrt{x_1^2-\cos^2\theta}-\sqrt{x_2^2-\cos^2\theta}+\sqrt{x_3^2-\cos^2\theta}.\nonumber
\end{eqnarray}
It is clear that
\begin{eqnarray}
f(\frac{\pi}{2})&=&1-x_1-x_2-x_3\leq 0,\\
g(\frac{\pi}{2})&=&1+x_3-x_1-x_2\geq 0,\\
f(\cos^{-1}(x_3))&=&g(\cos^{-1}(x_3))\in \mathcal{R},
\end{eqnarray}
from which we have
\begin{equation}
f(\frac{\pi}{2}).g(\frac{\pi}{2})\leq 0~\mbox{and}
~f(\cos^{-1}(x_3)).g(\cos^{-1}(x_3))\geq 0.
\end{equation}
Noticing that $f(\theta)g(\theta)$ is a real-valued function of
$\theta$, we conclude by the intermediate value theorem that there
exists $\cos^{-1}(x_3)\leq \theta_0\leq \frac{\pi}{2}$ such that
$f(\theta_{0})g(\theta_0)=0$. So $f(\theta_0)=0$ or $g(\theta_0)=0$.
Without loss of generality, assume $g(\theta_0)=0$, then we can
choose $u_{11}$, $u_{21}$, $u_{31}$ as follows:
\begin{eqnarray}
u_{11}=\sqrt{\frac{\sin\theta_0-\sqrt{x_1^2-\cos^2\theta_0}}{2\sin\theta_0}},\\
u_{21}=\sqrt{\frac{\sin\theta_0-\sqrt{x_2^2-\cos^2\theta_0}}{2\sin\theta_0}},\\
u_{31}=\sqrt{\frac{\sin\theta_0+\sqrt{x_2^2-\cos^2\theta_0}}{2\sin\theta_0}}.
\end{eqnarray}
By extending $(u_{11},u_{21}, u_{31})$ into an real orthogonal
matrix we obtain other entries $u_{kl}$. With that we complete the
proof of $\mathcal{P}_2\subseteq \mathcal{P}_1$.\hfill
$\blacksquare$

We strongly believe that it also holds $\mathcal{P}_1\subseteq
\mathcal{P}_2$, thus $\mathcal{P}_1=\mathcal{P}_2$. However, we
don't know how to give a rigorous proof for this up to now.

\smallskip

\section*{Acknowledgement}

We thank Z.-F. Ji, G.-M. Wang, J.-X. Chen, Z.-H. Wei, and C. Zhang
for helpful conversations. R. Duan acknowledges J. Walgate for
interesting discussions at QIP 2007.


\begin{thebibliography}{99}
\bibitem{NOTE1} It was intrdouced in Ref. \cite{BDM+99} that  a set of orthogonal product
states $\mathcal{S}$ is completable if it can be extended into an
orthogonal product basis of $\mathcal{H}$ (or local extension
$\mathcal{H}_{ext}=\otimes_{k=1}^K(\mathcal{H}_k\oplus
\mathcal{H}_k')$ of $\mathcal{H}$). Here we generalize this notion
by allowing $\mathcal{S}$ is an arbitary set of orthogonal separable
projecors such that the projector on the orthogonal complement of
$\mathcal{S}$ is again separable.
\end{thebibliography}
\end{document}